\documentclass[conference]{IEEEtran}
\IEEEoverridecommandlockouts

\usepackage{cite}
\usepackage{amsmath,amssymb,amsfonts}
\usepackage{algorithmic}
\usepackage{graphicx}
\usepackage{textcomp}
\usepackage{xcolor}
\usepackage{comment}
\usepackage{subcaption}
\def\BibTeX{{\rm B\kern-.05em{\sc i\kern-.025em b}\kern-.08em
    T\kern-.1667em\lower.7ex\hbox{E}\kern-.125emX}}

\begin{document}

	\title{Unleashing Sensor-Aided Environment Awareness for Beam Management in Beyond-5G Networks: An OpenAirInterface Experimental Platform\thanks{This work has received funding by the German Federal Ministry of Education and Research (BMBF) in the course of the 6GEM research hub under grant number 16KISK036K.}}

	\author{ \IEEEauthorblockN{
			Aron Schott \IEEEauthorrefmark{1},
			Berk Acikgöz \IEEEauthorrefmark{1},
			Omar Massoud \IEEEauthorrefmark{1},
			Marina Petrova \IEEEauthorrefmark{1}\IEEEauthorrefmark{2},
			Ljiljana Simi\'{c} \IEEEauthorrefmark{1}
		}
		
		\IEEEauthorblockA{
			\IEEEauthorrefmark{1}
			Institute for Networked Systems, RWTH Aachen University\\
			\IEEEauthorrefmark{2}
			Mobile Communications and Computing, RWTH Aachen University
		}
		
		\IEEEauthorblockA{
			\IEEEauthorrefmark{1}
			\{asc, bac, oma, lsi\}@inets.rwth-aachen.de\\
			\IEEEauthorrefmark{2}
			petrova@mcc.rwth-aachen.de
		}
	}

	\maketitle

	\begin{abstract}
		Large antenna arrays and beamforming techniques are key components to exploit the the spectrum-rich FR2 bands in next-generation mobile communication networks. Given the site-specific spatio-temporal variations of the mm-wave channel, non-RF sensor inputs and environment awareness can be leveraged to greatly enhance beam management decisions, e.g. via machine learning (ML) techniques. However, the current literature lacks open platforms to gather datasets for the training of such ML techniques and to evaluate novel beam management approaches in real-time, real-world scenarios and full-stack end-to-end networks. In this work, we present our SDR-based experimental platform based on OpenAirInterface and are the first to integrate popular low-cost antenna array transceivers, beam sweeping capabilities, and a highly-modular sensor framework and associated interfaces into such a full-stack experimental platform. This beam management experimentation in real-world, real-time scenarios and allows gathering datasets towards developing ML-based beam management protocols that incorporate environment awareness via multiple sensor modalities.		
	\end{abstract}


	\section{Introduction}
	\label{sec:introduction}
	Large antenna arrays enable exploiting higher frequency bands with large and previously unused bandwidth in beyond-5G cellular networks~\cite{mm_wave_data_rate}. However, the increase in pathloss, penetration loss, and susceptibility to blockage in the spectrum-rich FR2 (and high FR3) bands necessitates highly directional signal transmission through beamforming techniques~\cite{mm_wave_propagation}. To overcome the effect of beam misalignment through user equipment (UE) mobility and blockage caused by urban dynamics, agile beam management requires obtaining high-dimensional, directional channel state information (CSI), imposing a significant signaling overhead~\cite{mm_wave_delay}. Early commercial mm-wave deployments in 5G-NR have suffered from the poor real-world performance of such RF-based beam management protocols, thus failing to live up to the promises of mm-wave in 5G and highlighting the need to develop more sophisticated beam management approaches~\cite{verizon, SECON_2022}.
	
	To overcome this beam management burden towards enabling widespread exploitation of the mm-wave bands, leveraging sensor-aided real-time environment awareness through non-RF sensor inputs and machine learning (ML) is key to unlocking the potential of spectrum-rich bands for next-generation networks~\cite{ROY2023109729}. Towards this, several research groups have presented early experimental platforms and datasets for the development of novel beam management protocols, e.g.~\cite{Deepsense, mMobile, mmORAN}. However, these lack support for relevant 3GPP mm-wave frequency bands~\cite{Deepsense}, consider directional antenna arrays only to a limited extent~\cite{mMobile}, and do not integrate any sensor modalities~\cite{mmORAN}. Namely, an open platform that enables gathering a large number of open and comparable datasets in order to design and train ML-based environment-aware beam management approaches and assessing the real-world performance of novel sensor-aided beam management protocols is missing from the current literature.
		
	Importantly, evaluating the performance of such novel beam management approaches in \textit{full-stack end-to-end} networks is crucial to reveal system-level performance aspects, e.g. the control overhead for channel estimation~\cite{7744807} or the performance impact on higher protocol layers~\cite{8335489}. However, the majority of existing open mm-wave platforms, e.g.~\cite{WiNTECH, Sivers_3}, do not enable experiments in full-stack end-to-end networks. Moreover, the few existing datasets that have been gathered in commercial mm-wave network deployments~\cite{beam_management_commercial_2} and platforms that integrate some sensor input in full-stack networks~\cite{OAI_sensor} provide only limited or no insights into UE and next-generation Node B (gNB) beam management operations and observables, thus failing to provide a complete picture of the operation and performance of beam management in practice.	
				
	To address this, we present our mm-wave experimental platform which is built on OpenAirInterface, an open-source 5G-NR FR2-compliant network stack, and \textbf{(1)} integrate popular 28/60~GHz antenna array transceivers from Sivers Semiconductors to provide a low-cost alternative to existing solutions, \textbf{(2)} implement crucial beam management functions to enable real-time, real-world experimentation of novel beam management protocols, and \textbf{(3)} integrate multi-modal sensor inputs for beyond-5G radio resource management. By doing (1), we enable the broader community to leverage mm-wave research in OpenAirInterface with their already existing investments into the order-of-magnitude cheaper Sivers antenna array transceivers compared to other competitors like~\cite{TMYTEK}. With (2) and (3), we are the first to enable beam sweeping through the antenna array transceiver control directly implemented in OpenAirInterface and the first to provide a flexible framework and interfaces with OpenAirInterface to leverage sensor modalities. Thus, our experimental platform enables the design, real-time evaluation, and ML dataset generation for novel beam management and radio resource management approaches incorporating sensor information in real-world scenarios and full-stack end-to-end mm-wave networks. We plan to merge our implementation with the official OpenAirInterface code and provide a stable version in~\cite{github_link}.
		
	The rest of the paper is organized as follows. Sec.~\ref{sec:related_work} discusses the related work and Sec.~\ref{sec:platform} presents our experimental platform and the implementation details. Sec.~\ref{sec:bm} presents the verification of our beam management implementation and Sec.~\ref{sec:ml} demonstrates sensor-aided beam management using our experimental platform. Sec.~\ref{sec:conclusion} concludes the paper.

	\section{Related Work}
	\label{sec:related_work}
		
	The design and evaluation of novel beam management solutions in FR2 requires dedicated platforms for researchers to validate their approaches in real-time and in real-world scenarios. Furthermore, developing practical sensor-aided beam management solutions requires a large number of open and comparable datasets for ML training. Importantly, operating these experimental platforms and generating datasets in end-to-end full-stack networks and obtaining access to the underlying network parameters and procedures is crucial to reveal and incorporate system-level performance aspects that cannot be obtained otherwise, e.g. the realistic signaling overhead.~\cite{7744807}.	
		
	Over the past years, a number of mm-wave testbeds have been developed to facilitate real-world experimentation on beam management in mm-wave networks~\cite{COSMOS, mMobile, MIMORPH, HELIX, mmORAN, Deepsense, FLASH, WiNTECH, OAI_sensor}. However, these have important limitations: \cite{Deepsense, FLASH, MIMORPH} are restricted to 60~GHz antenna array transceivers, thus lacking support for the important 28~GHz frequency band; \cite{Deepsense, mMobile, FLASH} feature directional antenna arrays only on either the reciever (RX) or transmitter (TX); \cite{COSMOS, MIMORPH, HELIX} are reliant on inflexible and expensive FPGA implementations; and \cite{COSMOS, mMobile, MIMORPH, HELIX, mmORAN} do not integrate any sensor modalities to support sensor-aided beam management experimentation for beyond-5G networks. While~\cite{Deepsense, FLASH, WiNTECH, OAI_sensor} do integrate multi-modal sensor inputs, \cite{Deepsense, FLASH, WiNTECH} lack any network stack implementation. A recent exception is the work in~\cite{OAI_sensor}, demonstrating camera-aided beam switching and ML-driven object detection in OpenAirInterface. Importantly however, the work in~\cite{OAI_sensor} does not integrate beam switching directly in OpenAirInterface and relies on a commercial UE, thus limiting access and insights into the UE-side beam management operation. To address these limitations, our experimental platform integrates sensor modalities through a highly modular block design and relies solely on commercial-of-the-shelf components making our system reasonably affordable, highly customizable (in contrast to~\cite{COSMOS, MIMORPH, HELIX}), and easily extendable with additional sensors/RF front-ends. We build our experimental platform on top of and directly implement beam management into OpenAirInterface (in contrast to~\cite{OAI_sensor}), enabling the design and evaluation of novel beam management protocols in full-stack end-to-end networks.
	
	Several research groups have worked on the evaluation of existing commercial mm-wave deployments, e.g.~\cite{beam_management_commercial_2} and acquiring multi-modal datasets for the design of novel beam management approaches~\cite{Deepsense, FLASH}. The measurement campaign in~\cite{beam_management_commercial_2} found near-optimal beam selection in commercial 5G-NR deployments at the expense of high signaling overhead. However, due to commercial cellular networks being ''black boxes'', works like~\cite{beam_management_commercial_2} rely solely on limited UE-side observables, thus failing to provide full insights into the operation and performance of beam management in practice. While the authors in~\cite{Deepsense, FLASH} have developed novel sensor-aided beam management approaches based on their multi-modal datasets, none of the existing large-scale mm-wave datasets on environment awareness have been gathered in full-stack end-to-end networks, thus failing to provide insights into the system-level performance of such approaches. In contrast to~\cite{Deepsense, FLASH, OAI_sensor}, we build our experimental platform on top of the OpenAirInterface gNB and UE, enabling to generate full-stack mm-wave multi-modal sensor datasets with access to all relevant network parameters.
							
	The open-source OpenAirInterface 5G RAN project~\cite{OAI} has emerged as one of the main candidates for 3GPP-compliant 4G and 5G-NR cellular network research in the past years. To this end, there has been limited work on integrating mm-wave transceivers into OpenAirInterface, notably~\cite{mmORAN, TMYTEK}. While the authors in~\cite{mmORAN} utilize an unspecified and closed-source mm-wave phased antenna array, the solution from TMYTEK~\cite{TMYTEK} has gained increasing attention in the research community after the recent demonstrations of their FR2 antenna array transceiver integration into OpenAirInterface at conferences and tradeshows. However, TMYTEK does not yet integrate beam management into OpenAirInterface directly, thus not enabling the evaluation of standard-compliant and novel beam management approaches. Importantly, OpenAirInterface itself does not yet implement any hardware control of beam sweeping using antenna array transceivers. To address these limitations, we integrate the popular low-cost (i.e. compared to the TMYTEK solution) 28/60~GHz phased antenna array transceivers~\cite{SiversEVK} from Sivers into OpenAirInterface to provide an open-source alternative to the existing solutions. The majority of prior experimental works on mm-wave beam management and testbeds has been based on these antenna array transceivers~\cite{Deepsense, Sivers_3, Sivers_4, Sivers_5, Sivers_7}, making their integration into OpenAirInterface highly valuable for the research community. Finally, we integrate a generalized sensor interface into OpenAirInterface allowing beam management to leverage sensor modalities and detail one such example solution as demonstrated at the Berlin 6G Conference 2025. 
					
	\begin{figure}[t]
		\centering
		\subfloat[Experimental platform hardware setup and interfaces.\label{fig:HW_setup}]{\includegraphics[width=.98\columnwidth]{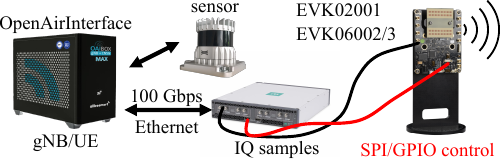}} \\
		\vspace{2pt}
		\subfloat[Flexible software framework based on Docker and ROS2.\label{fig:SW_framework}]{\includegraphics[width=.98\columnwidth]{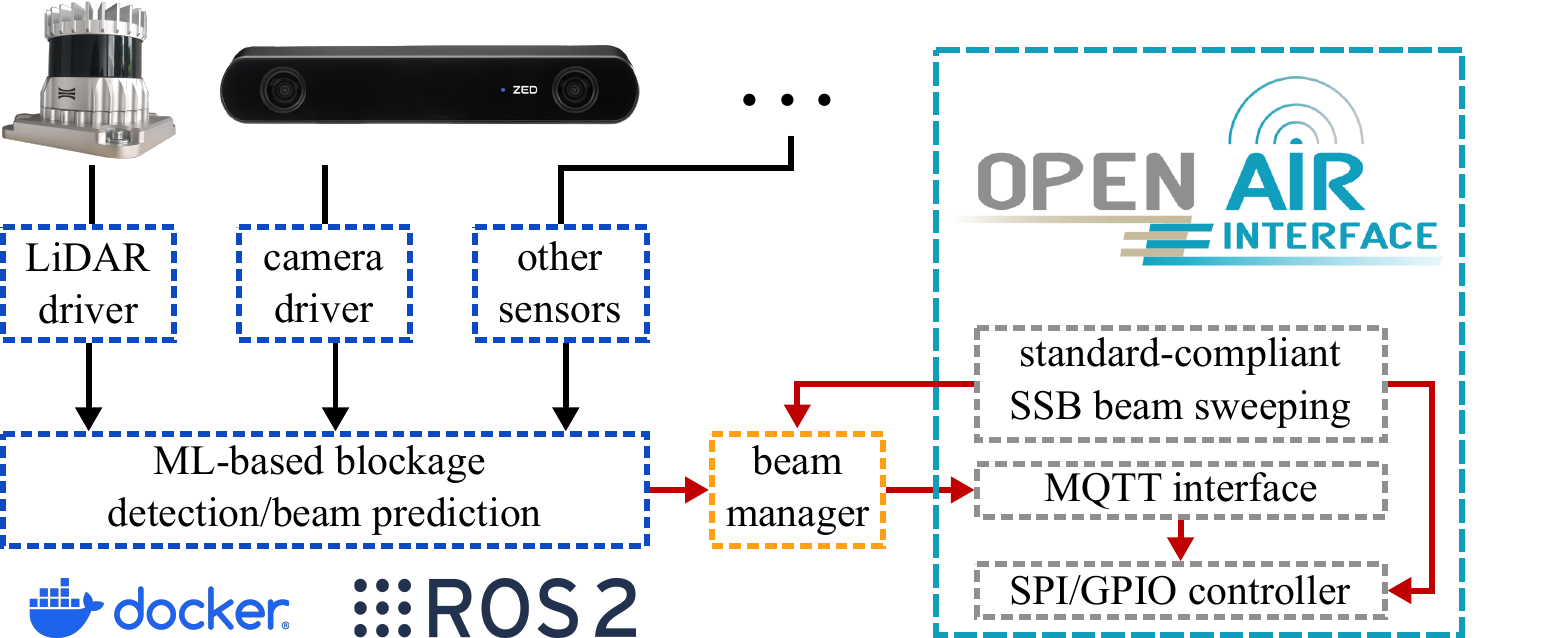}} \\
		\vspace{2pt}
		\subfloat[Fully-assembled experimental platform with gNB and UE.\label{fig:berlin}]{\includegraphics[width=.98\columnwidth, trim=0pt 120pt 0pt 95pt, clip]{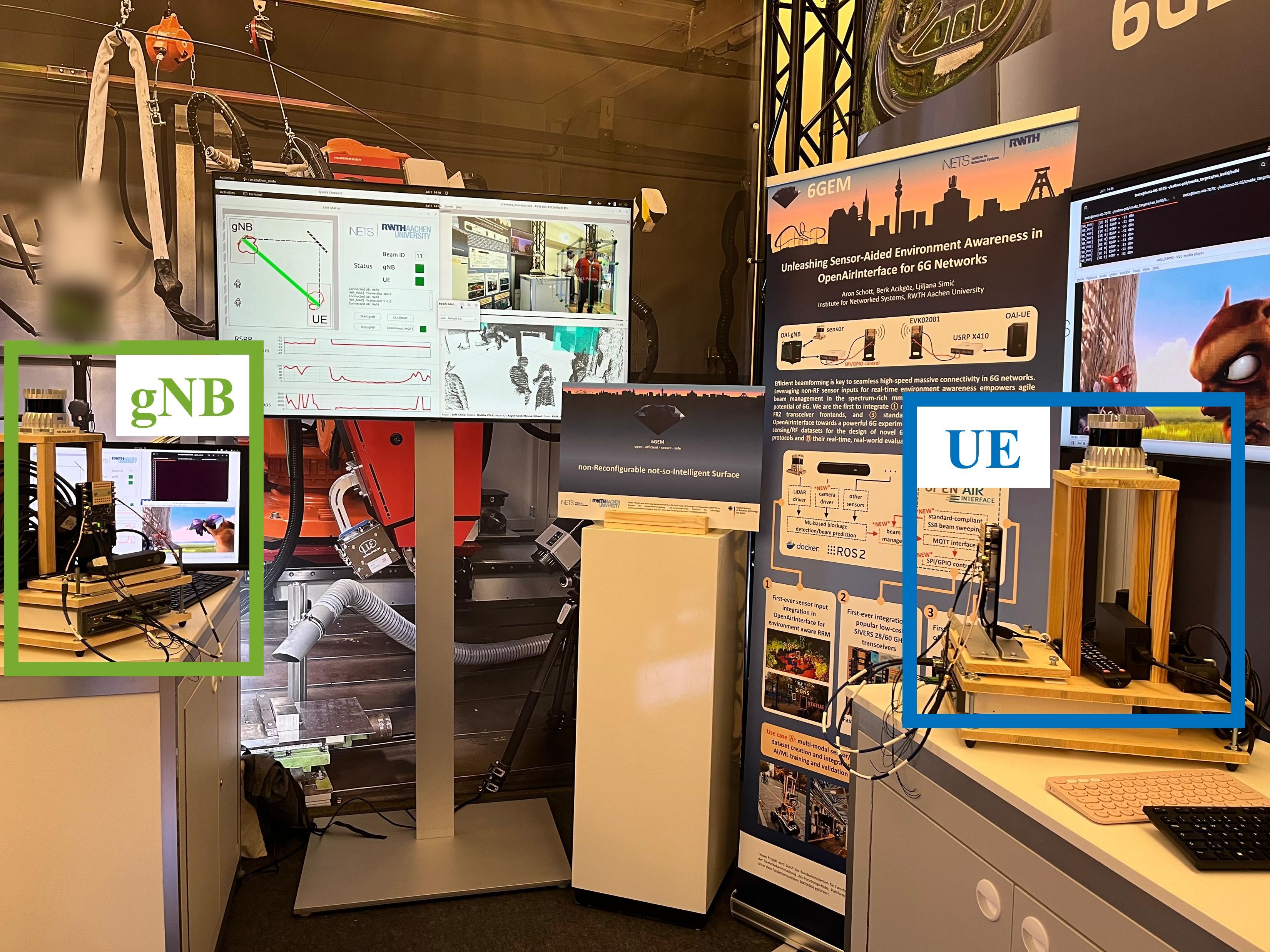}} \\
		\caption{Experimental platform architecture with (a) hardware components and setup, (b) software framework and interfaces, and (c) demonstration setup at the Berlin 6G Conference 2025.}
		\label{fig:platform}
	\end{figure}
			
	\section{Experimental Platform: Architecture \& Implementation}
	\label{sec:platform}
	In this section, we present our experimental platform towards enabling environment awareness and beam management experimentation in future full-stack mm-wave networks. The architecture of our experimental platform is shown in Fig.~\ref{fig:platform}, illustrating key hardware and software components and interfaces. We build the experimental platform on top of OpenAirInterface, the SDR USRP X410, and the Sivers direct up/down-conversion antenna array transceivers. We use Docker containers and the Robot Operating System 2 (ROS2) to implement a flexible block-based structure and interface it to OpenAirInterface via the MQTT protocol. To enable the reproduction of our experimental platform, Sec.~\ref{sec:sivers} details our antenna array transceiver integration, Sec.~\ref{sec:ssb} introduces our beam management implementation, and Sec.~\ref{sec:sensors} details our sensor framework and integration in OpenAirInterface.
	
	\subsection{Mm-Wave Antenna Array Integration}
	\label{sec:sivers}
	We are the first to integrate the popular low-cost 28~GHz (5G-NR FR2) and 60~GHz (5G NR-U) EVK02001 and EVK06002/3 phased antenna array transceivers from Sivers Semiconductors into OpenAirInterface, providing a low-cost alternative to the (only other OpenAirInterface integrated) mm-wave antenna array from TMYTEK. To enable fast and standard-compliant beam switch timings of 5G-and-beyond mm-wave systems, we directly implement beam switching of the two antenna array transceivers via Serial Peripheral Interface (SPI) in OpenAirInterface using the USRP Hardware Driver (UHD)~\cite{UHD_github, GPIOapi} and the USRP X410 front-panel General Purpose Input/Output (GPIO) interface. 
	
	\begin{figure}[t]
		\centering
		\includegraphics[width=0.90\columnwidth]{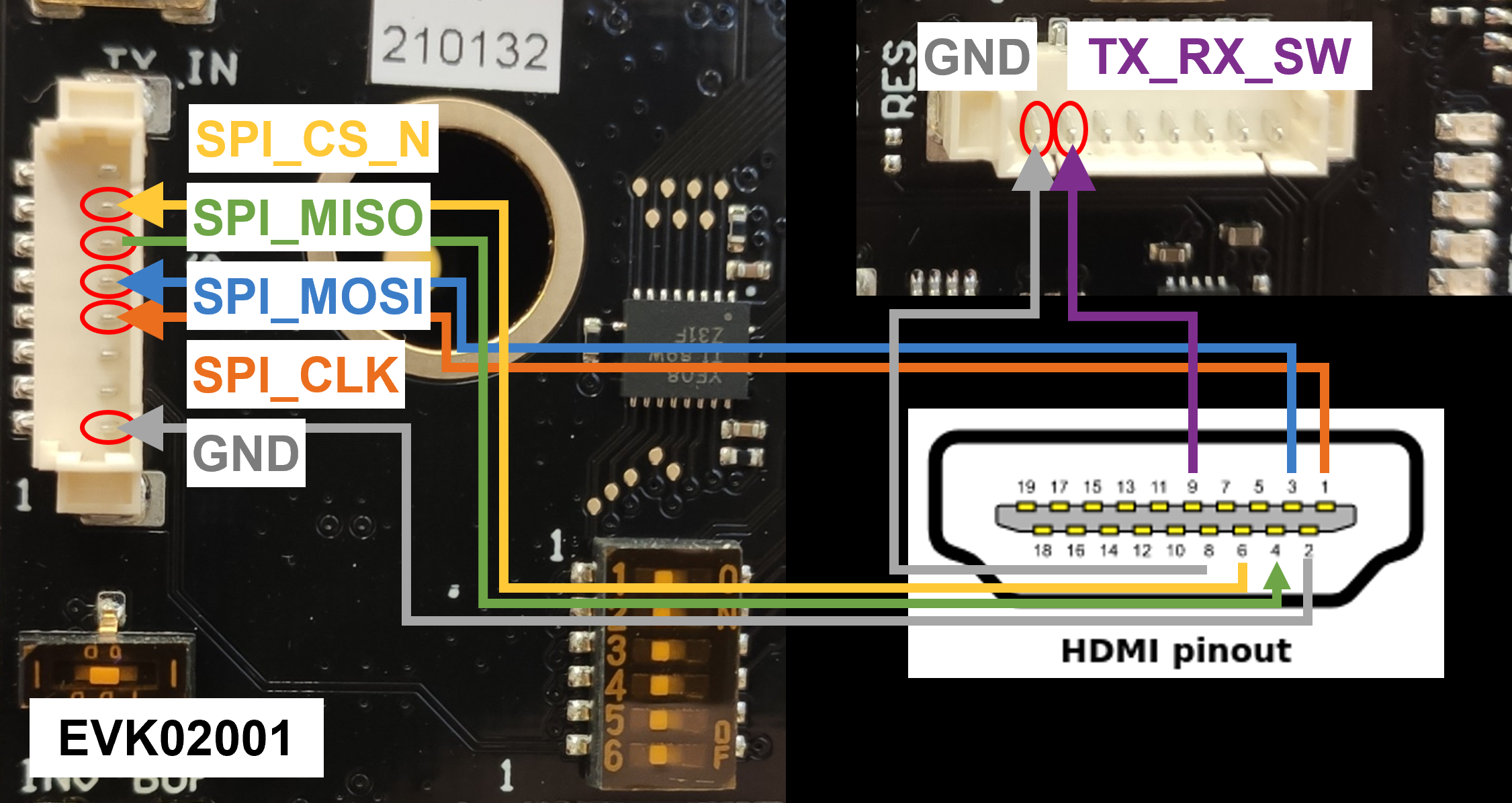}
		\caption{Pin connections between the Sivers phased antenna array transceiver EVK02001 and the USRP X410.}
		\label{fig:sivers_spi_gpio}
	\end{figure}	
				
	We implement a set of functions that handle antenna array initialization and beam switching according to the antenna array codebook beam ID. During initialization we specify the front-panel GPIO connector to be used for SPI operation, map the GPIO interface pins to the antenna array pins, and provide the SPI clock divider. For beam switching, we control the currently used beam using SPI write operations and two registers, namely \textit{bf\textunderscore tx\textunderscore awv\textunderscore ptr} and \textit{bf\textunderscore rx\textunderscore awv\textunderscore ptr} for TX and RX, respectively. 
	In order to meet the required TX/RX switch timings with our antenna array transceivers, we utilize the Automatic Transmit/Receive (ATR) registers of the USRP X410 for standard-compliant TX/RX switching. The antenna array transceivers require five GPIO pins and two ground connections for SPI beam switch and GPIO TX/RX switch operation: \textbf{SPI\textunderscore{CLK}} for timing/synchronization; \textbf{SPI\textunderscore{MOSI}} and \textbf{SPI\textunderscore{MISO}} for data transmission from and to the USRP; \textbf{SPI\textunderscore{CS\textunderscore{N}}} to identify the specific antenna array transceiver chip; and \textbf{TX\textunderscore{RX\textunderscore{SW}}} for TX/RX switching. Fig.~\ref{fig:sivers_spi_gpio} illustrates the connections between the HDMI breakout board plugged into the GPIO interface of the USRP X410 and the two \mbox{8-pin} cables with MOLEX terminals plugged into the EVK02001.
		
	\subsection{Multi-Beam SSB Implementation}
	\label{sec:ssb}	
	We are the first to implement and demonstrate 5G-NR-compliant codebook-based beam sweeping via Synchronization Signal Block (SSB) control signals and antenna array transceivers directly in OpenAirInterface. The software implementation of SSBs in OpenAirInterface has been ongoing since 2021. However, the beam switching procedures to the strongest beam identified through a beam sweep was only implemented in software in late 2024~\cite{beam_managemen_merge}. Importantly, executing beam sweeping in hardware, using antenna array transceivers, has not yet been implemented into OpenAirInterface. In fact, all current 5G-NR FR2 configurations for OpenAirInterface rely on a single fixed-beam SSB for initial access and eliminate beam sweeping entirely. Thus, prior platforms based on OpenAirInterface, e.g.~\cite{mmORAN}, do not enable real-time evaluation of beam sweeping in 5G-and-beyond mm-wave networks. To address this, we implement multi-SSB transmission for beam sweeping in OpenAirInterface and map the SSBs to the codebook-beams of antenna array transceivers via a new configuration file for the 5G-NR band n257 via the following variables:
	
	\begin{itemize}
		\item \textbf{ssb\textunderscore{PositionsInBurst\textunderscore{Bitmap}}} defines the SSB transmission pattern for up to 64 SSBs as a bitmap, where 0 and 1 indicate whether an SSB is disabled or enabled for transmission, respectively.
		\item \textbf{set\textunderscore{analog\textunderscore{beamforming}}} enables analog beamforming using codebook-based antenna array transceivers.
		\item \textbf{beam\textunderscore{weights}} maps the SSBs to the beam IDs of the antenna array transceiver codebook.
	\end{itemize}
	
	We verify SSB transmission and correct beam sweeping of the antenna array transceivers using our experimental platform and the 5G-NR downlink measurement application in a spectrum analyzer in Sec.~\ref{sec:bm}. Therefore, our experimental platform is a significant step towards enabling real-time studies of 3GPP-compliant downlink-based beam management and novel beam management protocols in real-world scenarios and full-stack end-to-end mm-wave networks. 

	\subsection{Sensor Integration}
	\label{sec:sensors}	
	We are the first to implement a generalized and flexible sensor interface into OpenAirInterface for sensor-assisted beam management	towards the development and testing of flexible sensor-aided environment-aware solutions in beyond-5G networks. The integration of sensor modalities in our experimental platform is based on Docker containers and ROS2, implementing the respective sensor drivers, ML-blocks for sensor data processing, a beam manager module, and an interface using MQTT to connect to OpenAirInterface allowing to fetch beam decisions derived from sensor modalities. This highly modular block design allows researchers to build their own experimental platform and adjust it to their needs. We note that our experimental platform currently targets beam management experimentation but the flexible design allows utilizing the platform in the more general sense for sensor-aided radio resource management and environment awareness in beyond-5G networks.
		
	Our experimental platform (\emph{cf.} Fig.~\ref{fig:platform}) features an Ouster OS-1 LiDAR with 128/2048 vertical/horizontal channels (200~m maximum range and 20~Hz frame rate) covering the environment in 360$^\circ$ azimuth and from -22.5$^\circ$ to 22.5$^\circ$ in elevation. A 120$^\circ$ field-of-view StereoLabs ZED 2i stereo camera is used to capture the environment for vision-based schemes and serves as a ground-truth reference for LiDAR post-processing. Both the LiDAR and camera drivers are based on ROS2, running within respective sensor driver Docker containers. We deploy an ML block that is based on the camera's built-in object detection framework to identify obstacles in the vicinity of the experimental platform. Alternatively, we can deploy an ML block that identifies objects within the LiDAR pointcloud. Both ML blocks feed information about obstacles to the beam manager block, which decides whether an active communication link is about to be blocked and what actions to be taken to avoid interruptions. Importantly, this beam manager block allows integration of custom sensor-aided beam management approaches and their real-time evaluation in real-world scenarios full-stack mm-wave networks.
	
	\begin{figure}[t]
		\centering
		\includegraphics[width=0.98\columnwidth]{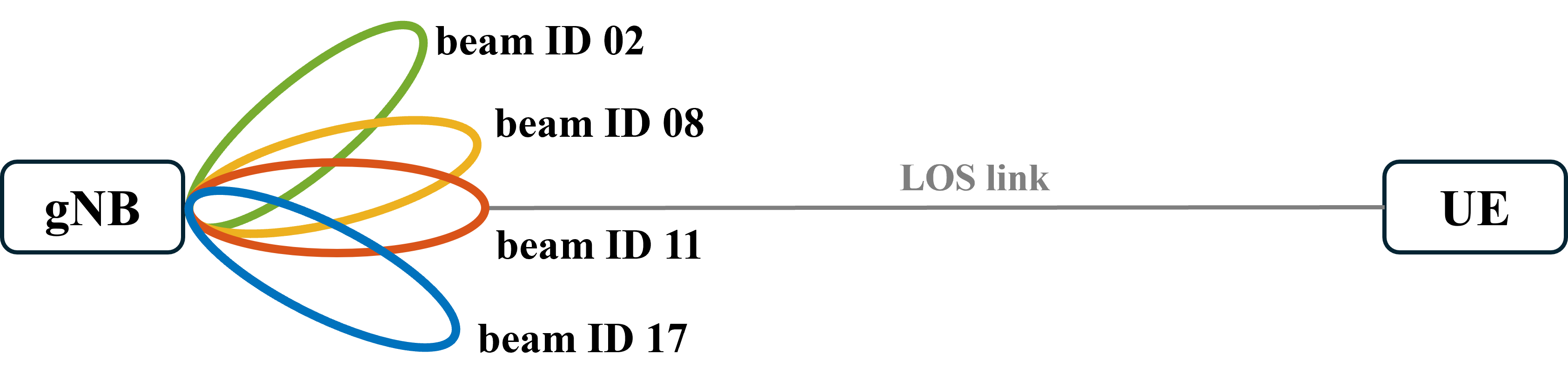}
		\caption{SSB beam sweeping validation setup with SSBs to be sent over four different antenna codebook beams on the gNB and the UE being fixed to an omnidirectional codebook beam.}
		\label{fig:scenarios}
	\end{figure}
		
	\section{Verification of 5G-NR Downlink Beam Management in OpenAirInterface}
	\label{sec:bm}	
	To validate the SSB beam sweeping capabilities implemented in OpenAirInterface, we set up our experimental platform in a test scenario using the gNB and UE in the 28~GHz frequency band over a distance of 4~m. To the best of our knowledge, this is the first time that beam sweeping using both the OpenAirInterface gNB and OpenAirInterface UE has been demonstrated. We connect the antenna array transceiver on the UE to a FSW67 spectrum analyzer running the 5G-NR downlink measurement application to verify standard compliance of our implementation. The gNB is configured to transmit/receive at an intermediate frequency of $f_{IF}=533.28$~MHz and a bandwidth of $B=50$~MHz, which is then directly up/downconverted to the carrier frequency of $f_c=27.533$~GHz by the antenna array transceivers. For our test, we enable four SSBs to be sent over four different antenna codebook beams on the gNB (\emph{cf.} Fig.~\ref{fig:scenarios}). We note that we can support up to 64 antenna codebook beams in compliance with the 5G-NR standard but limited the number of beams used in the test scenario due do current stability issues of the OpenAirInterface gNB and UE code stack.
	
	\begin{figure}[t]
		\centering
		\subfloat[5G-NR downlink measurement application of a spectrum analyzer with four SSBs visible in the time/frequency domain and successful synchronization of the application with the OAI gNB.\label{fig:SSBs}]{\includegraphics[width=.98\columnwidth, trim=0pt 55pt 0pt 76pt, clip]{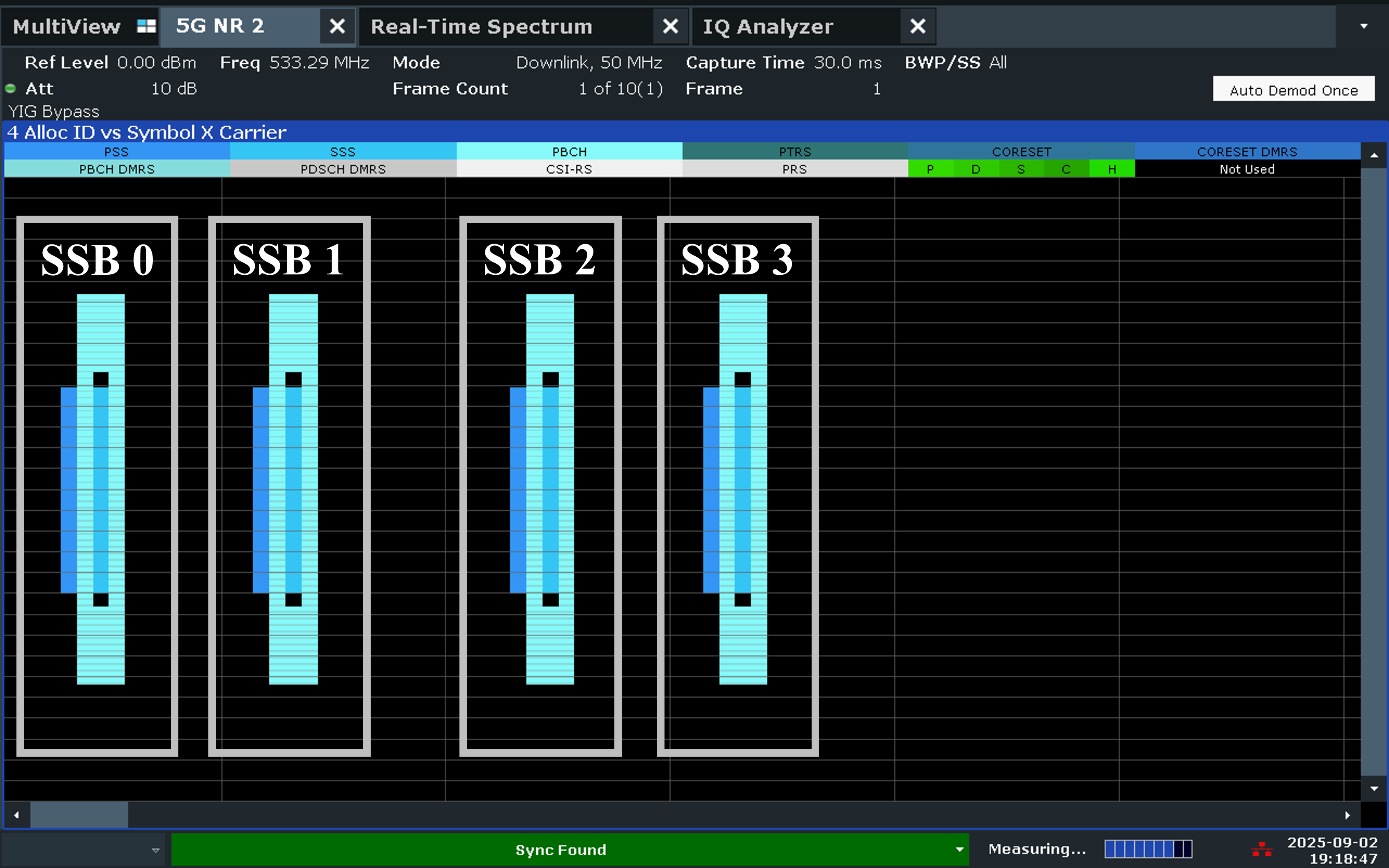}} \\
		\vspace{2pt}
		\subfloat[Spectrum analyzer screenshot of recorded IQ samples from SSB transmission over time for the fixed beam (ID 11) scenario.\label{fig:SSB_IQ_same}]{\includegraphics[width=.98\columnwidth, trim=0pt 213pt 0pt 65pt, clip]{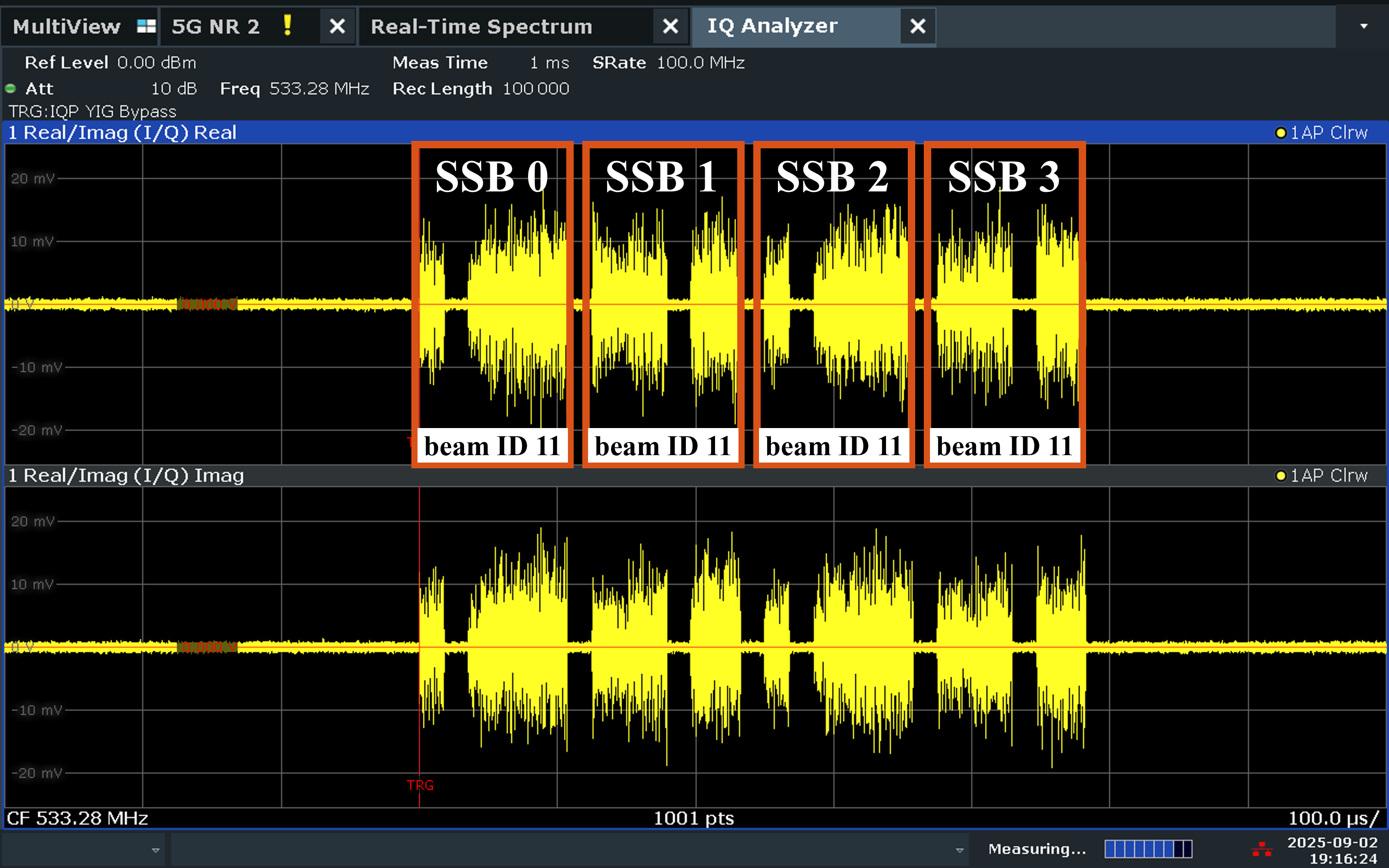}} \\
		\vspace{2pt}
		\subfloat[Spectrum analyzer screenshot of recorded IQ samples from SSB transmission over time for the beam sweeping scenario.\label{fig:SSB_IQ_diff}]{\includegraphics[width=.98\columnwidth, trim=0pt 213pt 0pt 65pt, clip]{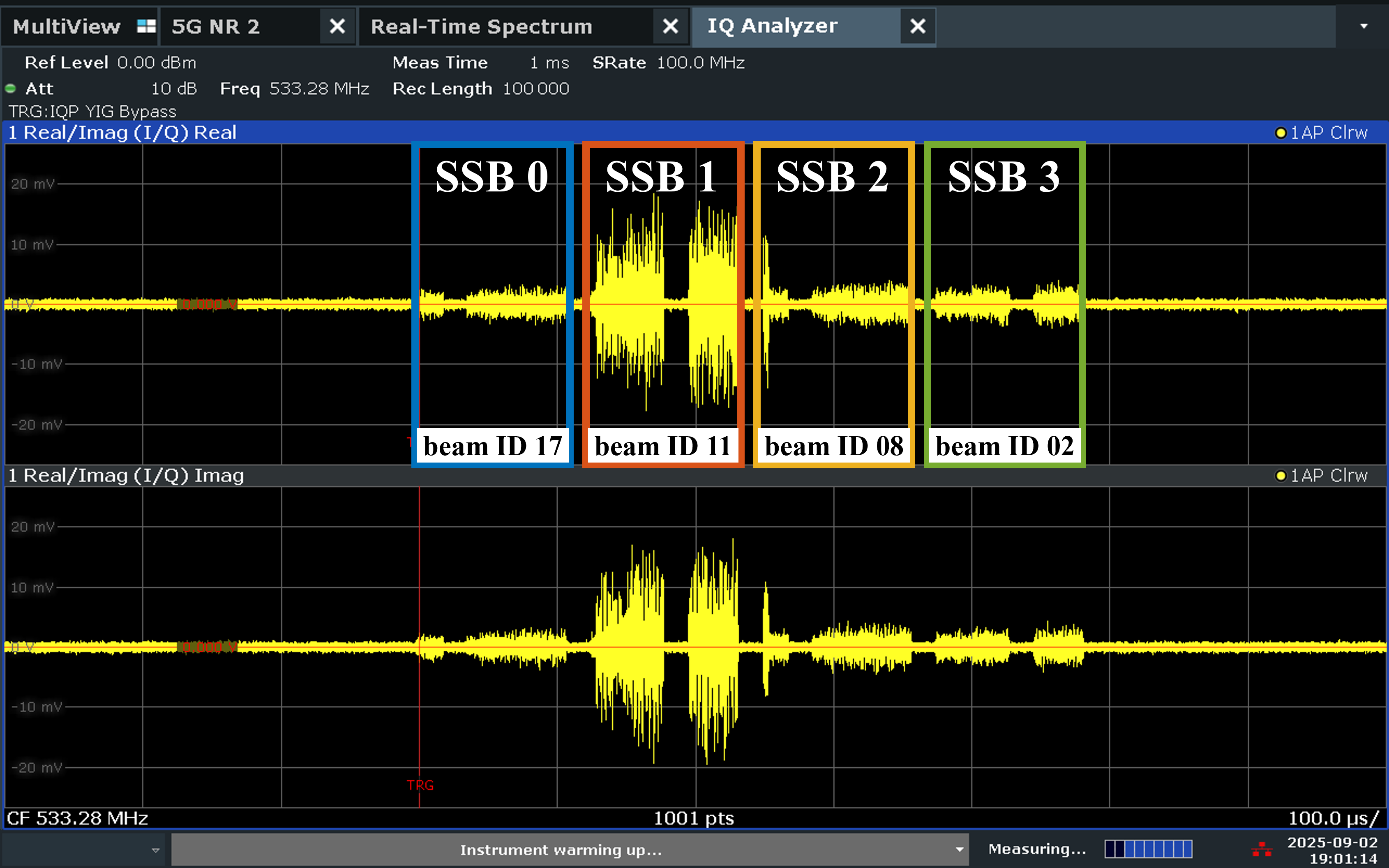}} \\
		\caption{Verification of SSB transmission in (a) time/frequency domain and the corresponding IQ samples over time for (b) the fixed beam and (c) the beam sweeping scenario.}
		\label{fig:SSB_verification}
	\end{figure}
		
	We consider the following scenarios: \textbf{(1)} multi-SSB transmission without beam sweeping to verify the correct scheduling and transmission of SSBs for a fixed codebook beam and \textbf{(2)} multi-SSB transmission with beam sweeping to validate the antenna array transceiver correctly iterating through the beam codebook. Fig.~\ref{fig:SSBs} shows the 5G-NR downlink measurement application of the spectrum analyzer for scenario 1 with the four SSBs clearly visible in the time/frequency domain. Fig.~\ref{fig:SSB_IQ_same} shows the reception of these four SSBs as IQ samples over time. Due to the antenna array transceiver not changing its codebook beam during this scenario, the four SSBs are received with a comparable signal strength in Fig.~\ref{fig:SSB_IQ_same}. In contrast Fig.~\ref{fig:SSB_IQ_diff} shows the received SSB IQ samples for scenario 2, i.e. the antenna array transceiver on the gNB iterating through it's codebook beam according to our pre-defined SSB-to-codebook beam ID mapping in the OpenAirInterface gNB configuration file. The second SSB in Fig.~\ref{fig:SSB_IQ_diff} shows the highest signal strength due to this SSB using the line-of-sight (LOS) beam 11 between the gNB and UE. In contrast, the other SSBs can be observed at a significantly lower signal strength due to the gNB-UE beam misalignment during SSB transmission. With our antenna array transceivers and codebook-based beam switching capabilities directly integrated in OpenAirInterface, our experimental platform enables standard-compliant and cost-effective beam management experimentation in full-stack mm-wave networks and real-time real-world scenarios.
	
	\begin{figure}[t]
		\centering
		\includegraphics[width=0.98\columnwidth]{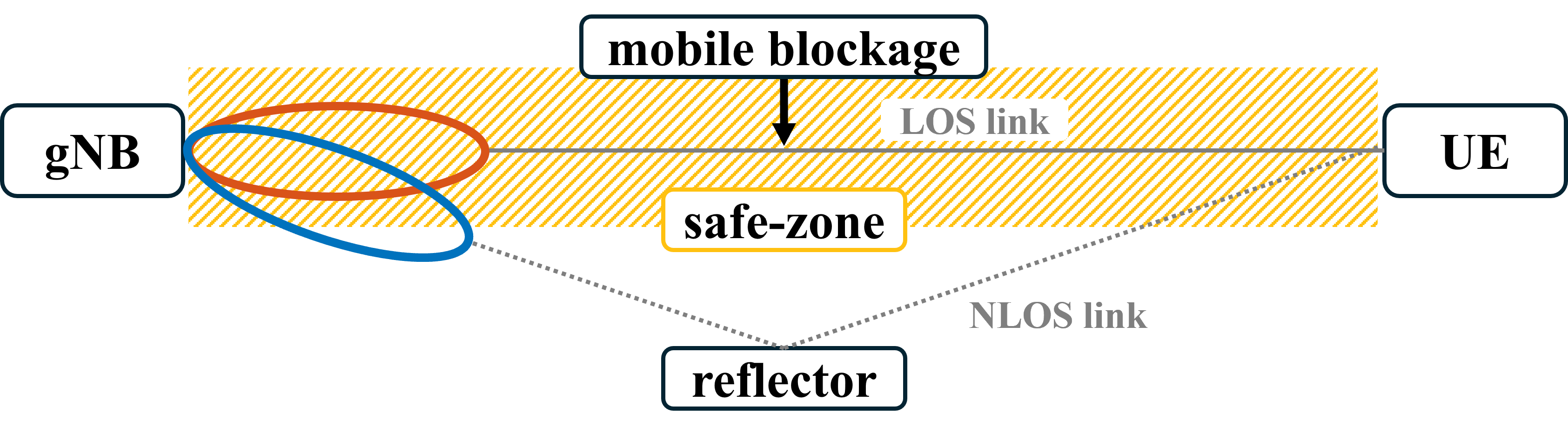}
		\caption{Demonstration setup with LOS link, NLOS link via a reflector, mobile blockage, and the pre-defined safe-zone for camera-aided beam management.}
		\label{fig:safezone}
	\end{figure}
			
	\section{Demonstration of sensor-aided beam management in OpenAirInterface}
	\label{sec:ml}	
	Integrating multi-modal sensor inputs in full-stack mobile communication networks enables the unique opportunity to evaluate novel environment-aware beam management solutions in real-time and in real-world environments. We demonstrated one such solution at the Berlin 6G Conference 2025 with a live audience (\emph{cf.} Fig.~\ref{fig:berlin}), incorporating camera and LiDAR sensor data to assist in beam management decisions and a live video stream to highlight the real-time performance of sensor-aided beam management.
	
	The demo setup is illustrated in Fig.~\ref{fig:safezone}, with a gNB-UE separation of 3~m, a reflective surface ensuring a secondary non-LOS (NLOS) link, the UE using an omnidirectional beam, and the gNB using two different beams for either the LOS or NLOS communication link. After establishing the connection between the UE and gNB over the LOS link, pedestrians acting as mobile blockages were able to repeatedly walk into this link and attempt to degrade the performance. For this demonstration, we defined a safe-zone stretching 1~m around the primary LOS communication link (\emph{cf.} Fig.~\ref{fig:safezone}) which enables proactive beam switching to the secondary NLOS link once an obstacle, detected by an ML-based object detection block processing the sensor data (\emph{cf.} Fig.~\ref{fig:SW_framework}), is entering this safe-zone in an attempt to block the primary LOS link.
	 		
	\begin{figure}[t]
		\centering
		\includegraphics[width=0.98\columnwidth]{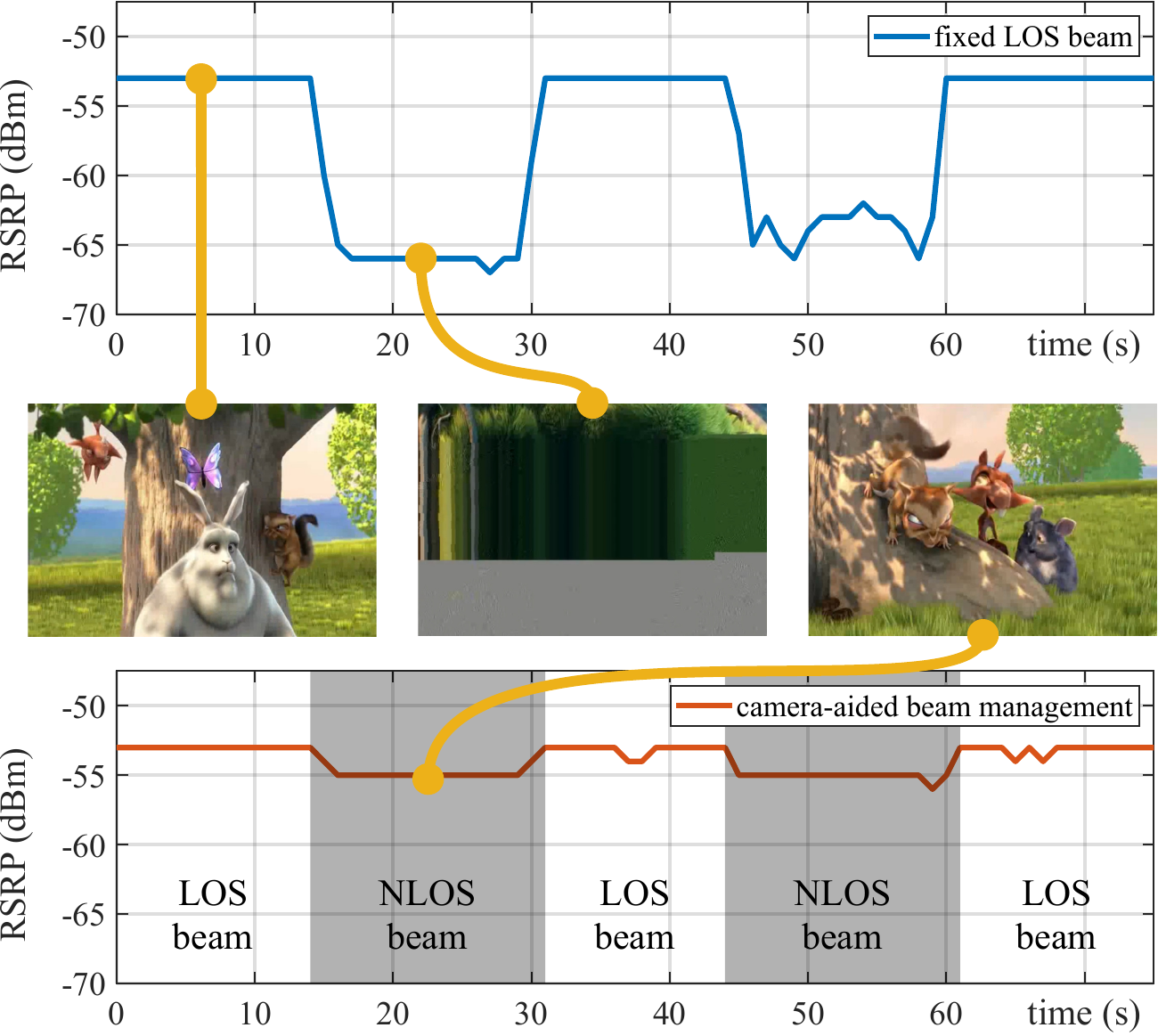}
		\caption{UE-experienced RSRP over time and screenshots of the over-the-air received video affected by the link quality for the fixed LOS beam scenario (top) and camera-aided beam management (bottom).}
		\label{fig:oai}
	\end{figure}

	Fig.~\ref{fig:oai} shows the UE-experienced reference signal received power (RSRP) over time and screenshots of the video stream received at the UE for the following two scenarios: \textbf{(1)} a fixed LOS beam and \textbf{(2)} camera-aided beam management, i.e. preemptively switching between the LOS and NLOS link in the presence of obstacles. With an unobstructed LOS link for $t\in [0, 13]$, the UE experiences an RSRP of -53~dBm in both scenarios, resulting in a stable video stream. For the fixed LOS beam scenario, the UE-experienced RSRP significantly drops to \mbox{-66~dBm}, e.g. at $t=22$~s, due to a pedestrian blocking the LOS link and causing the disrupted video stream. Once the pedestrian leaves the LOS link, the UE-experienced RSRP and the video stream recover. In contrast, our camera-aided beam management scenario in Fig.~\ref{fig:oai} shows barely any performance degradation in RSRP and no degradation of the live video stream, thanks to the camera and the ML-based object detection identifying the pedestrian breaching the safe zone and the beam manager issuing a beam switch decision to OpenAirInterface to switch to the secondary NLOS link prior to the pedestrian blocking the link. By doing so, we demonstrate the effectiveness of proactive beam management procedures that incorporate environment-awareness to deliver seamless connectivity in mm-wave networks. This serves as an example of how OpenAirInterface with environment awareness can be used directly to develop and experimentally evaluate novel sensor-aided beam management protocols.
	
	\section{Conclusions}
	\label{sec:conclusion}
	We presented our experimental platform towards environment awareness for beam management in beyond-5G networks, integrating the popular Sivers antenna array transceivers, enabling beam sweeping through the antenna array transceiver control directly implemented in OpenAirInterface, and implementing a generalized and highly modular sensor framework and associated interfaces. We validated our beam sweeping implementation via standard-compliant 5G-NR application running on a spectrum analyzer and demonstrated how sensor inputs in full-stack networks can assist in beam management decision making. Thus, our experimental platform enables the design, real-time evaluation, and ML dataset generation for novel sensor-aided beam management and radio resource management approaches in real-world scenarios and full-stack end-to-end mm-wave networks.	

	\bibliographystyle{IEEEtran}
	\bibliography{Bibliography.bib}
	
\end{document}